Title: **A low-power integrated magneto-optic modulator on silicon for cryogenic applications**

Authors: Paolo Pintus[*,1], Leonardo Ranzani[2], Sergio Pinna[1], Duanni Huang[1,3], Martin V. Gustafsson[2], Fotini Karinou[4], Giovanni Andrea Casula[5], Yuya Shoji[6], Yota Takamura[6], Tetsuya Mizumoto[6], Mohammad Soltani[2], John E. Bowers[1]

[*]Corresponding author: ppintus@ece.ucsb.edu

Affiliations:

[1]Electrical and Computer Engineering Department, University of California Santa Barbara, California 93106, USA

[2]Raytheon BBN Technologies, 10 Moulton Street, Cambridge, Massachusetts 02138, USA

[3]Intel Corporation, 2200 Mission College Blvd, Santa Clara, California 95054, USA

[4]Microsoft Research Ltd, 21 Station Road, Cambridge CB1 2FB, United Kingdom

[5]Department of Electrical and Electronic Engineering, University of Cagliari, Via Castelfidardo, Cagliari 93100, Italy

[6]Department of Electrical and Electronic Engineering, Tokyo Institute of Technology, 2-12-1 Ookayama, Meguro-ku, Tokyo, 152-8552, Japan

**Abstract**:

A fundamental challenge of the quantum revolution is to efficiently interface the quantum computing systems operating at cryogenic temperatures with room temperature electronics and media for high data-rate communication. Current approaches to control and readout of such cryogenic computing systems use electrical cables, which prevent scalability due to their large size, heat conduction, and limited bandwidth[1]. A more viable approach is to use optical fibers which allow high-capacity transmission and thermal isolation[2,3]. A key component in implementing photonic datalinks is a cryogenic optical modulator for converting data from the electrical to the optical domain at high speed and with low power consumption, while operating at temperatures of 4 K or lower. Cryogenic modulators based on the electro-optic effect have been demonstrated in a variety of material platforms[3–8], however they are voltage driven components while superconducting circuits are current based, resulting in a large impedance mismatch. Here, we present the first demonstration of an integrated current-driven modulator based on the magneto-optic effect operating over a wide temperature range that extends down to less than 4 K. The modulator works at data rates up to 2 Gbps with energy consumption below 4 pJ/bit, and we show that this figure can be reduced to less than 40 fJ/bit with optimized design and fabrication. This modulator is a hybrid device, where a current-driven magneto-optically active crystal (cerium substituted yttrium iron garnet, or Ce:YIG) is bonded to a high-quality silicon microring resonator. Because of its potential for extremely low power consumption under cryogenic conditions, the class of magneto-optical modulators demonstrated here has the potential to enable efficient data links in large-scale systems for quantum information processing.

**Manuscript**:

In emerging cryogenic classical and quantum computing systems, there is a need for transferring massive amounts of information from cryogenic circuitry to room temperature, while avoiding significant hardware complexity and heat load. This becomes particularly important as the systems scale up, considering the limited physical space and cooling power available in cryogenic systems. The realization of photonic integrated circuits (PICs) operating at low temperatures and the use of optical fibers to connect different temperature stages can successfully overcome those limitations, thus enabling scalable, low-cost, and power-efficient optical interconnections for large data transfer rates[1–3].

Silicon photonics is perhaps the most promising technology platform for the development of such cryogenic PIC components, due to its promising scalability, compatibility with CMOS and superconducting electronic manufacturing[9], and excellent optical performance. Despite many advantages of silicon photonics, low temperatures (e.g., ~< 4 K) pose significant limitations on conventional modulators and switches. Thermo-optic switches become non-functional due to a significant drop of the thermo-optic coefficient[10], and modulators based on plasma dispersion suffer from free-carrier freeze out which makes them inefficient at such low temperatures[11]. The carrier freeze out effect can be compensated by augmenting the doping concentration, but only at the cost of significantly increased optical absorption[8]. As a result, novel approaches for realizing energy efficient modulators and switches for cryogenic systems must be explored.

So far, only electro-optic modulators have been shown to work at low temperature[3–8], and the use of magneto-optic effects in integrated optics has been limited to optical isolators and circulators[12,13], with a few theoretical proposals existing for magneto-optical modulation[14–16]. This is largely due to the difficulties in manufacturing integrated magneto-optic devices, the challenges in applying fast time-variant magnetic fields, and the generally slower response of the magneto-optic effect compared to the electro-optic effect. In this work, we demonstrate the first high-speed magneto-optic modulator operating at temperatures as low as 4 K. This device is made by combining a magneto-optic garnet crystal with a silicon waveguide resonator and integrating an electromagnet to modulate the refractive index of the garnet. The proposed solution is particularly efficient at cryogenic temperatures, where the magneto-optic effect becomes stronger[17] and the power consumption in the electromagnet decreases drastically compared to room temperature due to reduced ohmic loss. The fabricated modulator shows an excess loss of only 2.3 dB, a modulation bandwidth up to 2 GHz with an energy consumption as low as 4 pJ per bit of transferred information, which can be reduced below 40 fJ/bit by optimizing the device design and fabrication process. This first demonstration of a current-driven magneto-optical modulator opens the path to further investigations on novel magneto-optic materials operating at low temperature to address the need for power-efficient data-link for the next generation of superconducting supercomputers and quantum computers[18–20].

The modulator demonstrated here is based on an all-pass silicon microring resonator with a cerium-substituted yttrium iron garnet (Ce:YIG) bonded on top, Fig. 1(a). The light that circulates in the microring resonator interacts evanescently with the garnet, and is affected by its magneto-optic properties. A gold micro-strip, aligned with the silicon microring, serves as an integrated electromagnet and is used to control the magneto-optical (MO) properties of the Ce:YIG, while a silicon optical waveguide is used to couple light in and out of the resonator.

The main idea of our work is to exploit the refractive index variation induced by an applied magnetic field to control the resonant wavelength of the microring, and hence to modulate the light passing through the waveguide. When an electrical current passes through the micro-strip, the Ce:YIG is magnetized by a transverse in-plane magnetic field, Fig. 1(b), changing the effective refractive index of the optical mode[21] (see the supplementary section for details). For the light in the microring, a variation of the effective index, $\Delta n_{eff}$, causes a proportional shift of the optical spectrum of the resonator

$$\Delta\lambda_{MO} = \lambda \frac{\Delta n_{eff}}{n_g} \qquad (1)$$

where $\lambda$ is the optical wavelength and $n_g$ is the group index of the optical mode[22]. If the resonant wavelength of the microring is aligned with the wavelength of the input signal, the light is fully attenuated and dissipates in the microring. When the direction of the current is reversed, the microring resonance tunes away from the incoming light which can then propagate through the waveguide unperturbed. Based on this concept, we designed and fabricated the device shown in Fig. 1(c).

Equation (1) is crucial for understanding the operating principle of this device, so we highlight a few important features: (i) reversing the current (i.e., the in-plane magnetic field in the Ce:YIG) changes the sign of $\Delta n_{eff}$; (ii) the effective index change differs for the clockwise and counter-clockwise propagating modes, having the same amplitude but opposite signs (nonreciprocal effect); (iii) the effective index change is largest when the optical mode is polarized along the y-axis (transverse magnetic polarized mode); (iv) the value of the effective index is proportional to the Faraday rotation constant (see the supplementary section for details) that is expected to increase with decreasing temperature[17,23].

The results of the modelling and device design are summarized in Fig. 2. The magnetic field distribution shown Fig. 2(a) is obtained after optimizing the cross-sections of the gold micro-strip (Fig. 2b) and the optical waveguide (Fig. 2c) to produce the largest shift in resonance wavelength for a given current. For the realization of this first prototype, we chose Ce:YIG as a magneto-optic material because it is optically transparent to telecom wavelengths and has one of the largest Faraday rotation constants in this wavelength range at room temperature. The optimized cross-section of the silicon waveguide is 600 nm wide and 220 nm tall, with a 400 nm thick Ce:YIG layer bonded on top[24]. To avoid bending loss, the diameter of the microring is chosen to be 70 µm. At room temperature, this cross-section guarantees $\Delta\lambda_{MO}$ = ±350 pm when the magnetization of the material is saturated in the two directions[24], i.e. $|H_x| \geq 5$ mT. A layer of 5 μm-thick substituted gadolinium gallium garnet (SGGG) covers the Ce:YIG and is a remnant of the substrate onto which the Ce:YIG was grown. On top of this layer, directly above the silicon microring, a gold electrode with a cross section of 3 μm × 1.5 μm. This is the electromagnet that modulates current that serves as a signal input port by generating a magnetic field through the Ce:YIG.

At 4K, the thermal shift of the microring resonance is negligible[10,25], and the current flowing through the electromagnet can be used to red-shift or blue-shift the resonance of the microring, depending on its direction. To perform amplitude modulation, the input laser is tuned a few tens of picometers away from the zero-current microring resonance wavelength, Fig. 2 (d). The current in the electromagnet is used to bring the resonance of the microring closer to or further away from the input wavelength. With an input current alternating between -110 mA and +110 mA and a microring with a quality factor of Q=25 000, we expect an extinction ratio (ER) of the modulated output signal larger than 10 dB.

The ER of the modulated signal depends on both the quality factor of the resonator and the amplitude of the current in the electromagnet. When the resonator is critically coupled to the waveguide[24] (that is, when the power coupling ratio between the microring and the waveguide, $K$, equals the intrinsic cavity loss during one roundtrip in the microring, γ), the ER can in principle be chosen to be arbitrarily large, limited only by fabrication variabilities. In Fig. 2(e), the ER of the modulated signal is computed as a function of the current amplitude for devices with different coupling conditions. The highest ER can be achieved at the critical coupling condition and when the current in the electromagnet is about 80 mA. Comparing over-coupled (K> γ) and under-coupled (K< γ) resonators with the same maximum ER, we notice that the under-coupled condition is more favorable for minimizing the driving current and hence the energy consumption.

A notable feature of this device architecture is the dramatic reduction in energy consumption for modulation at low temperature compared to room temperature. The average energy-per-bit (EPB) consumption is

$$\overline{E_b} = \frac{RI^2}{r_b} + \frac{1}{2}LI^2 \qquad (2)$$

where *R* and *L* are the effective resistance and inductance of the MO modulator, respectively, *I* is the modulation amplitude, and $r_b$ is the bit rate (see derivation in the supplementary material).

At low temperature, the resistance of the electromagnet drops, and can be made to vanish if the normal metal is replaced with a superconductor. In addition, we also observed that the magnetic dissipation is negligible in the Ce:YIG (see supplementary materials). As a result, the EPB depends only on the modulating current and the inductance of the device for a signal source with a suitable impedance. Assuming gold electrodes, the resistance drops by more than two orders of magnitude[26], going from 1.22 Ω at 300 K down to 12 mΩ at 4 K. On the other hand, the inductance does not change significantly with temperature and equals 0.25 nH for a 70 μm diameter coil. The average EPB for the proposed device is computed as a function of the modulating current amplitude and reported in Fig. 2(f) for T=300 K, 77 K and 4 K with $r_b$ = *1 Gbps*.

Combining the calculated values of the EPB (Fig. 2(g)) and the ER (Fig. 2(f)), we find that an ER as high as 10 dB can be achieved with less than 400 fJ/bit of dissipation for $r_b$ = *1 Gbps*. By thinning the SGGG layer further, the distance between the electromagnet and the optical waveguide can be reduced from 5 μm down to 1 μm, thus reducing the modulating current requirements by more than 50% and lowering the EPB to less than 100 fJ.

The magneto-optic modulator is fabricated using a fully silicon compatible process. The silicon microring resonator and the waveguide are patterned on a standard silicon-on-insulator (SOI) wafer with 220 nm of silicon on top of 2 μm of silicon dioxide ($SiO_2$). The Ce:YIG layer is grown on a separate

SGGG substrate in a process that yields high-quality crystals. Subsequently, Ce:YIG is bonded to the silicon wafer in a flip-chip process based on plasma activated $SiO_2$-$SiO_2$ covalent bonding at low-temperature[27]. After the bonding, the SGGG is mechanically polished in order to reduce the distance between the electromagnet, deposited on top, and the Ce:YIG/silicon interface (more details on the device fabrication can be found under "Methods" below). Using a straight silicon waveguide without Ce:YIG as a reference and comparing it to the modulator device when far removed from resonance, we estimate the excess loss due to the bonded layers to be no more than 2.3 dB at 1550 nm.

The performance of the device was tested at cryogenic temperatures in a closed-cycle cryostat (Montana Cryostation s200) at temperatures ranging between 4 K and 77 K. The thermo-optic (TO) resonance shift is determined from measurements of the microring spectral response when no current is applied to the electromagnet. Tracking the shift of the resonance with respect to the temperature for both the clockwise (CW) and counter-clockwise (CCW) resonant modes, as shown in Fig.3 (a), we extract the thermo-optic coefficient (see the supplementary section for more details)

$$\frac{dn_{eff}}{dT} = \frac{\lambda}{n_g}\frac{d\lambda}{dT} \qquad (3)$$

This is plotted in Fig. 3b together with computed TO contributions from silicon[5] and silica[25] for reference. At 4 K, a temperature increment of ΔT=1 K causes the resonance to shift by less than 0.05 pm, making the TO effect negligible for our applications. At higher temperatures, the TO shift increases, reaching 16.8 pm/K at 77 K and 80.0 pm/K at 300 K, respectively. In the temperature range 4 K to 77 K, we observe that the resonances in the two directions are not perfectly aligned, but their offset is a mere 3.7±1.9 pm. This suggests the presence of a very small residual in-plane magnetization in the Ce:YIG (see supplementary materials for more details).

When a current is injected in the electromagnet, the Ce:YIG is locally magnetized and the optical response is no longer reciprocal. That is to say, the effective indices for the CW and CCW propagation directions are different, resulting in the splitting of the resonant wavelength for the two directions ("MO split"). As shown in Fig. 3(c), the MO split changes linearly with the current amplitude at all temperatures under investigation: 300 K, 77 K and 4 K. Although the Faraday rotation of Ce:YIG is expected to increase for decreasing temperature[17,23], we find that the MO split at 4 K and 77 K is about six times lower than the corresponding value measured at 300 K for the same current (i.e., the same magnetic field). As reported in the supplementary material, the Ce:YIG results magnetically harder at lower temperature, and the weaker MO split is owing to the stronger coercive force measured at 77 K and 4 K compared to the room temperature one.

The spectral response of the modulator is insensitive to the thermal fluctuations produced by the gold electromagnet, as shown in Fig. 3(d). The large red-shift of the spectrum observed at room temperature drops significantly at 77K and becomes negligible at 4K. This is due to the enhanced conductivity of gold at cryogenic temperature, leading to lower dissipation, long with the previously discussed reduction in TO with decreasing temperature.

We evaluate the performance of the modulator at cryogenic temperature by measuring its bandwidth and its modulation fidelity when driven with high-speed data. For the first measurement, we use a tunable laser to generate an optical carrier at a wavelength near 1550 nm, which couples in and out of the device via lensed fibers and is collected after modulation by a high-speed photodetector at room temperature. The RF input signal is generated by a vector network analyzer and swept from 100 MHz to 20 GHz (more details on the experimental set-up are provided in the supplementary material). The measured frequency response of the magneto-optic modulator is plotted in Fig.4(a), showing a bandwidth as large as 6 GHz at a temperature of 4 K. The maximum modulation bandwidth is limited by the magnetic response of the Ce:YIG, which is on the order of hundreds of picoseconds[14]

To further characterize the bandwidth of the device, we performed data modulation measurements by sending a pseudo-random non-return-to-zero (NRZ) bit sequence of length $2^9-1$ to the modulator with a maximum modulation amplitude of 110 mA. To achieve a better signal-to-noise ratio, the modulated optical output signal is amplified by an erbium-doped fiber amplifier (EDFA) and filtered before reaching the photodetector, which is connected to a digital oscilloscope. For this measurement, the carrier wavelength is set close to 1550 nm. Although the ER around 1550 nm is low compared to shorter wavelengths (Fig. 4(b) and Fig. 4(c)), and the resonator is slightly over-coupled at 1550 nm ($K$ = 0.11 and $\gamma$ = 0.063), this wavelength offers the benefit of a high optical gain in the EDFA. The recorded modulation eye diagrams for 1 Gbps and 2 Gbps are shown in Fig. 4(d) at both 77 K and 4 K. Recalling the results shown in Fig. 3(c), we need six times higher current at these temperatures than at room temperature to achieve the same MO split. On the other hand, the smaller MO split in the Ce:YIG is compensated by the lower electrical resistivity in metals at cryogenic temperatures, which means that a larger current can be injected in the device without increasing the power consumption. In the fabricated device, we measure an input resistance of 1.43 Ω at 300 K, which drops to ~350 mΩ around 77 K, where it saturates. The low-temperature resistance is limited by the quality of the gold film. Improving the gold deposition conditions can lead to a resistance of ~250 mΩ at 77 K, and a reduction by up to another factor 20 at 4 K, with an expected value as low as 12 mΩ (see supplementary material).

The low energy consumption is one of the major benefits of the cryogenic magneto-optic modulator. As shown in Fig. 4(d), we experimentally measured an eye diagram up to 2 Gbps with an energy consumption as low as 3.9 pJ/bit at 4 K. Nonetheless, there is vast room for improvement by optimizing the modulator design and fabrication. In the device under test (DUT), the value of $E_b$ is limited by the resistance of the electromagnet, as shown in Fig. 5(a). The amount of the dissipated power can be reduced by improving the quality of the electromagnet or replacing it with a superconducting magnet, leading to $E_b \approx LI^2/2$ at 4 K, as shown in Fig. 5(b). At this point, the EPB can be further lowered by diminishing the inductance of the electromagnet and decreasing the modulating current. In the tested device, the measured value of L is 0.3 nH, which can be reduced to 0.2 nH in considering a microring with a diameter of 40 μm. Additionally, the current required for modulation can be reduced in two ways: (i) by considering an under-coupled resonator with the same ER, which yields a reduction in current of 50 % as shown in Fig.2(g); (ii) by decreasing the distance between the electromagnet and the silicon/Ce:YIG interface from 5.5 μm to 1.5 μm, which results in a reduction in required current amplitude by up to another factor 2.25 (see supplementary material). With such feasible design improvements, the energy per bit of the proposed device can be reduced by more than 30 times, as shown in Fig. 5(c), thus achieving an energy consumption as low as 40 fJ/bit.

In this work, we have experimentally demonstrated a new class of high-speed optical modulator based on the magneto-optic effect for cryogenic applications. The device is fully compatible with the silicon photonics platform, has a compact footprint of 70 × 70 μm$^2$, and an excess loss of 2.3 dB. The magnetic field that provides the optical modulation can be effectively controlled by an integrated electromagnet. At temperatures as low as 4 K, the energy consumption is below 4 pJ/bit for $r_b$ = 2Gbps, and has the potential to be greatly reduced by reducing the device footprint, thinning the SGGG substrate, and using under-coupled resonators. Such improvements will allow us to reduce the modulating current by a factor of 30, leading the $E_b$ below 40 fJ/bit.

The proposed cryogenic device is the first demonstration of an integrated current-driven optical modulator, which makes it seamlessly integrable with superconducting circuitry for quantum and cryogenic applications. The stronger coercive force measured at low temperatures and the maximum time response of the YIG, which is about 100 ps[14], are two limiting factors for the minimum power consumption and the maximum modulation bandwidth. Nevertheless, the promising results achieved in this early realization ought to lead the research towards magneto-optic materials with improved performance at cryogenic temperatures, which is still an unexplored area of investigation.

## Methods:

### Device Fabrication

A 220 nm silicon-on-insulator (SOI) wafer with 1 µm of buried oxide was patterned using a 248 nm ASML 5500 DUV stepper, and dry etched using a Bosch process (Plasma-Therm 770) to form the waveguides and resonators. In preparation for wafer bonding, both the SOI and the Ce:YIG/SGGG sample are rigorously cleaned, and activated with $O_2$ plasma (EVG 810). The Ce:YIG is directly bonded onto the SOI patterns using a flip-chip bonder (Finetech), and then annealed at 200°C for 6 hours under 3MPa of pressure to strengthen the bond. The required alignment accuracy is fairly tolerant (~200 µm). After the bonding, a 1 µm layer of SiO2 is sputtered everywhere on the chip as an upper cladding. Next, the SGGG substrate is thinned by mounting the sample against a flat chuck, and polishing (Allied Technologies) using a series of increasingly fine lapping films. The thickness of the SGGG is monitored using a micrometer and confirmed to be ~5 µm with a separate Dektak profilomety measurement. Variation of the thickness across the sample is roughly ±1.5 µm due to imperfect leveling of the chuck. The patterns for the metal microstrips and contacts are defined on the backside of the SGGG with a 365nm GCA i-line wafer stepper. Then, 22 nm of Ti is deposited as an underlayer, followed by 1.5µm of Au using electron-beam evaporation, and the metal microstrips and contacts are released with a lift-off procedure. Finally, the sample is diced and the facets are polished.

### Acknowledgements

This material is based upon work supported by the Air Force Office of Scientific Research under award number FA9550-21-1-0042. Any opinions, findings, and conclusions or recommendations expressed in this material are those of the authors and do not necessarily reflect the views of the United States Air Force. The authors also acknowledge Microsoft Research for supporting this research. The authors would like thank Paul Morton and Jon Peters for the useful discussions.

**Author Contributions**

P.P conceived and designed the device, performed the modelling, and analyzed the performance. P.P., L.R., S.P., and M.G. performed the DC and RF measurements at cryogenic temperatures. D.H. performed the mask layout and fabricated the device. P.P. and A.C. performed the RF simulations. P.P., L.R., S.P., and F.K. performed the analysis of the energy consumption. Y.S., Y.T., and T.M. grew the Ce:YIG samples and provide the magneto-optic material characterization, P.P., M.S. and J.E.B. supervised and coordinated the project. All authors contributed to the preparation of the manuscript.

**Additional Information**

The authors declare no competing financial interests. All data generated and analyzed during this study are included in this article and its supplementary information files. Data are available from the corresponding author upon request.

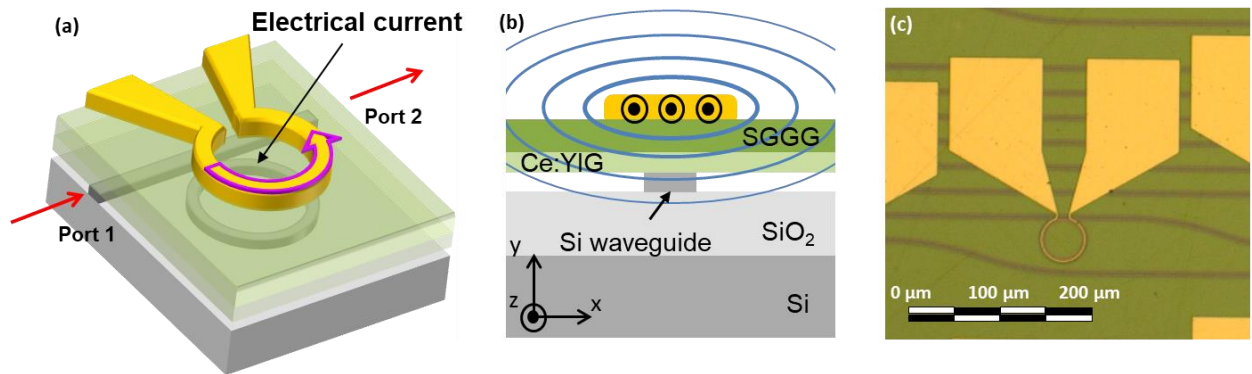

**Figure 1: Integrated magneto-optic modulator**. (a) Perspective view of the device (not to scale). The top gold coil is used to apply a radial magnetic field that magnetizes the Ce:YIG underneath. The silicon microring and the silicon waveguide, in the all-pass filter configuration, are visible through the transparent top-cladding. (b) Cross-section of the microring and electromagnet (not to scale) where the direction of the electrical current and the magnetic field are highlighted. (c) Optical micrograph of the fabricated sample (top view).

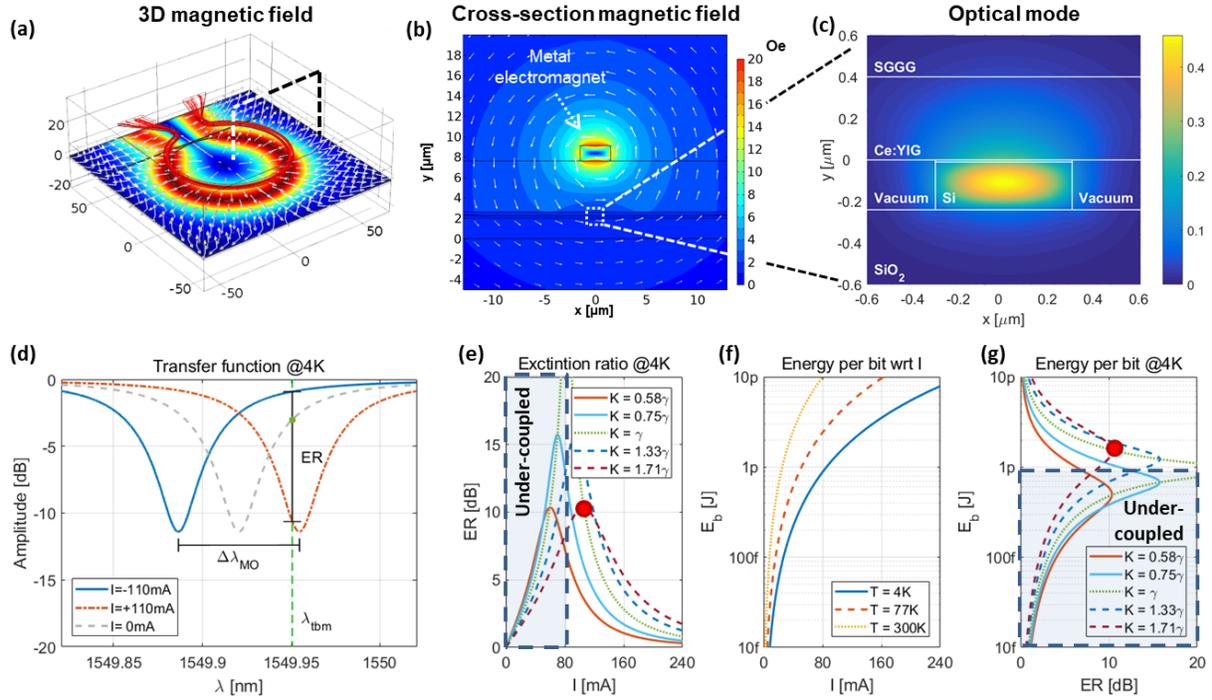

**Figure 2: Cryogenic magneto-optic modulator, design and optimization. (a)** Three-dimensional simulation of the modulating magnetic field generated by the electric current in the metal microstrip. The streamlines of the current are highlighted in red while the intensity of the in-plane radial magnetic field is shown in the Ce:YIG plane. On the same plane, the arrows indicate the direction of the magnetic field. **(b)** Magnetic field distribution in the device cross-section when a current of 10 mA is flowing through the electromagnet. **(c)** Cross-section mode profile, where a silica layer of 10 nm is assumed between the silicon microring and the bonded Ce:YIG layer. **(d)** The red-shifted and the blue-shifted spectral response of the device are shown when the current in the electromagnet is +110 mA (dashed pointed red curve) and -110 mA (continuous blue curve), respectively. As a reference, the spectral response of the microring modulator when no current is injected in the electromagnet (dashed gray curve) is also plotted. These curves refer to a resonator with Q = 25 000. The input laser wavelength is shown ($\lambda_{tbm}$, green dashed line) along with the extinction ratio (ER) and the MO split ($\Delta\lambda_{MO}$). **(e)** ER as a function of the modulating current for several microring-based modulators. The dot refers to the device we measured, where the coupling coefficient (*K=0.11*) and the round-trip loss (*γ=0.063*) have been computed from the measured ER and full-width at half-maximum of the spectral response. **(f)** Energy consumption per bit as a function of the modulating current at different temperatures. The increasing conductivity of gold with decreasing temperature results in a net improvement under cryogenic operation. In case of superconductor magnet, the power dissipated in the resistance vanishes. **(g)** Energy per bit as a function of ER for microring-based modulator in under-coupled, critically coupled and over-coupled conditions. The dot refers to the device under test.

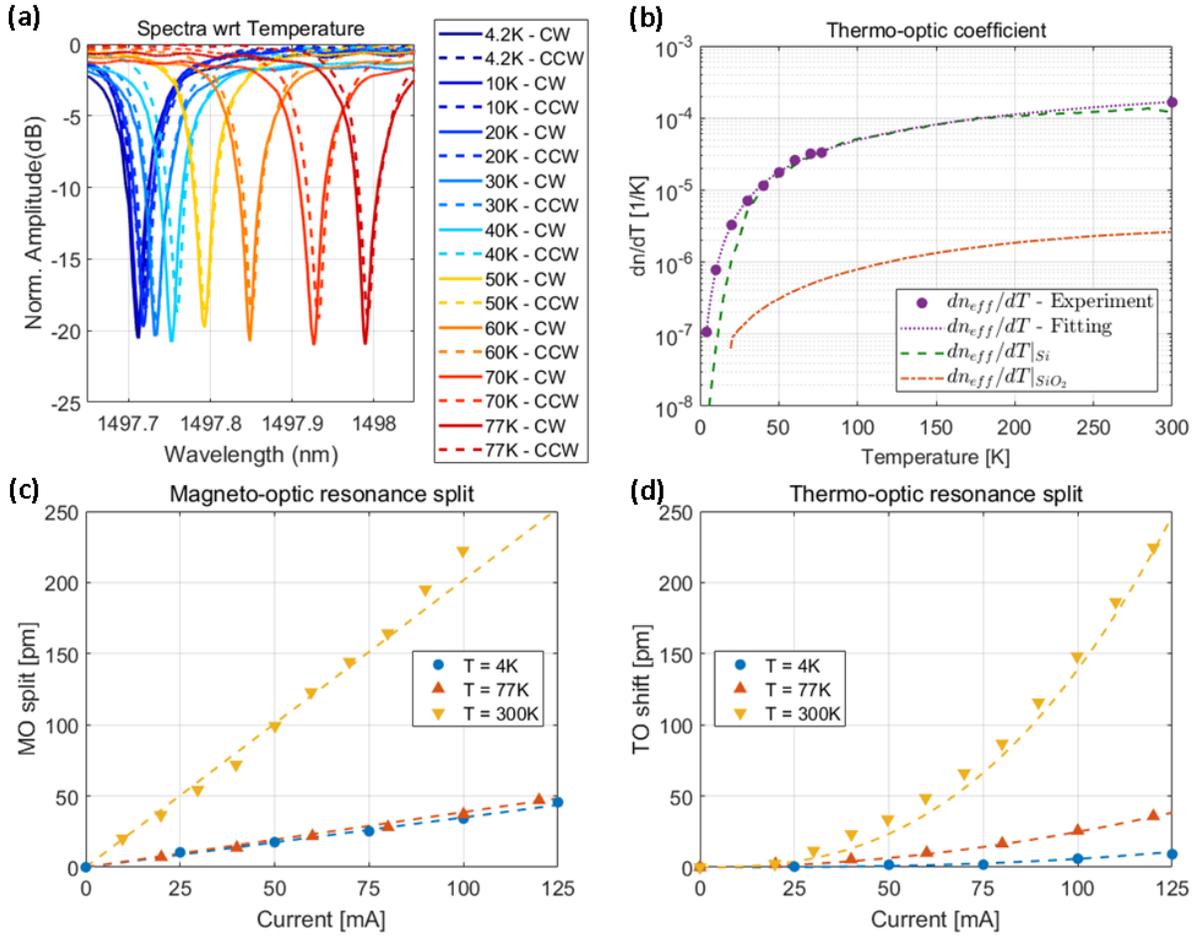

**Figure 3. Characterization of the thermo-optic (TO) and magneto-optic (MO) response of the modulator at cryogenic temperature**. **(a)** Thermo-optic shift of the spectral response of the microring between 4 K and 77 K when no current is injected in the electromagnet. The spectra for both clockwise (CW, solid lines) and counter-clockwise (CCW, dashed lines) propagation spectra are shown. **(b)** Thermo-optic coefficient of the device extracted from the thermal shift of the spectra. As a reference, the thermo-optic contribution of the silica and silicon are also shown. **(c)** The MO split of the microring resonance when a current is injected in the electromagnet. The CW and the CCW resonance split increases linearly as a function of the current magnitude at all temperatures. **(d)** TO shift of the microring resonance when a current is injected in the electromagnet because of Joule local heating. As a guide for the eye, the experimental results have been fitted with a sub-quadratic curve (~$I^b$ where b<2) since the conductivity of gold decreases when a current is injected in the electromagnet (increasing temperature) and the thermo-optic coefficient is not constant (Fig.2(b)).

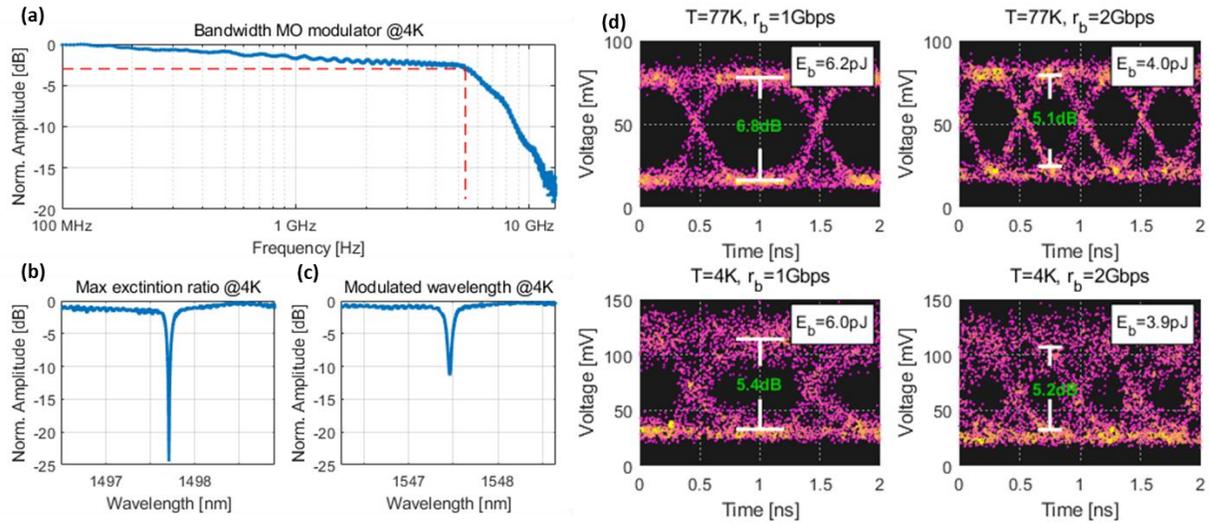

**Figure 4. High-speed characterization of magneto-optic modulator at cryogenic temperatures.** **(a)** Frequency response (Bode diagram) of the magneto-optic modulator at 4 K measured with a vector network analyzer. **(b)** Spectral response of the microring resonator around 1500 nm **(c)** Spectral response of the microring resonator around 1550 nm. **(d)** Eye diagram at bit rates of 1 Gbps (first column) and 2 Gbps (second column) measured at temperatures of 77 K (first row) and 4 K (second row). The eye diagram at 4 K shows a smaller ER compared to 77 K, which is mainly due to the difficulties in finely control the fiber-to-chip alignment. At temperatures around 4K, XYZ piezoelectric nano-positioners tend to "freeze" such that higher voltage is required, diminishing the alignment resolution (see supplementary material for more details on the measurement set-up). In the inset, we report the energy consumption per bit extracted from the measurements. The consumption per bit drops at low temperature due to the higher conductance of the electromagnet.

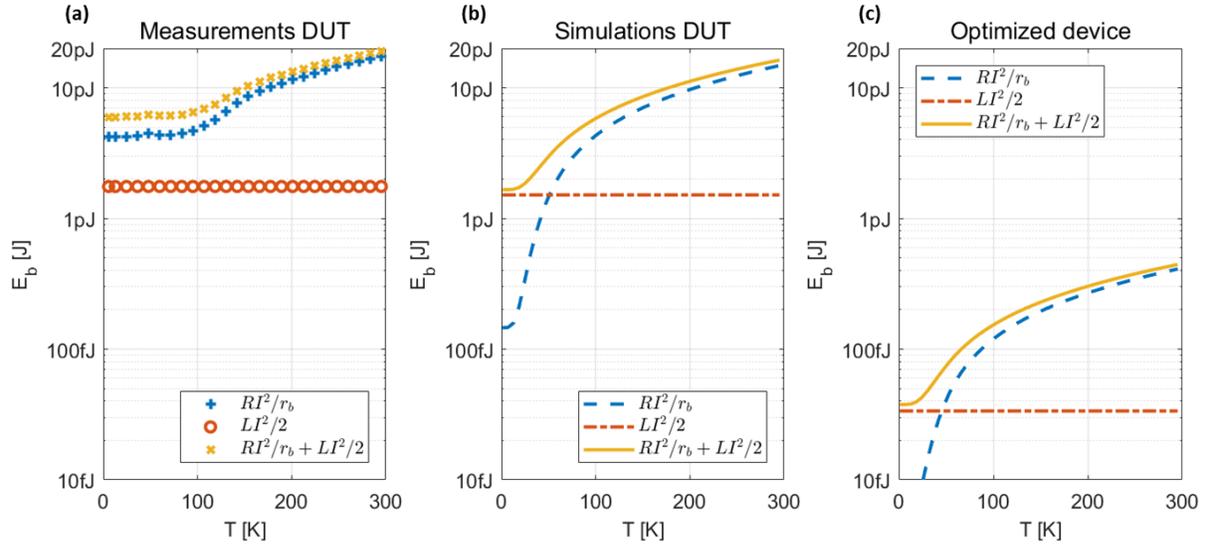

**Figure 5. Energy-per-bit of the magneto-optic modulator as a function of temperature for a bit rate of 2 Gbps. (a)** Measured energy-per-bit of the device-under-test (DUT). The measured value of $E_b$ is as low as 3.9 pJ/bit at 4 K and it is limited by the resistance of the electromagnet. **(b)** Simulated energy-per-bit of the device-under-test (DUT). A comparison between simulations and measurements shows strong agreement for the term $LI^2/2$. On the other hand, the theoretical value of $RI^2/r_b$ shows room for improvement by reducing the cryogenic resistance of the electromagnet. **(c)** Energy-per-bit that is achievable after optimizing the device by reducing the footprint, thinning the SGGG substrate, and considering under-coupled resonators.

Supplementary material of
# A low-power integrated magneto-optic modulator on silicon for cryogenic applications

Paolo Pintus[1], Leonardo Ranzani[2], Sergio Pinna[1], Duanni Huang[1,3], Martin V. Gustafsson[2], Fotini Karinou[4], Andrea Casula[5], Yuya Shoji[6], Yota Takamura[6], Tetsuya Mizumoto[6], Mohammad Soltani[2], John E. Bowers[1]

Affiliations:
[1]Electrical and Computer Engineering Department, University of California Santa Barbara, California 93106, USA
[2]Raytheon BBN Technologies, 10 Moulton Street, Cambridge, Massachusetts 02138, USA
[3]Intel Corporation, 2200 Mission College Blvd, Santa Clara, California 95054, USA
[4]Microsoft Research Ltd, 21 Station Road, Cambridge CB1 2FB, United Kingdom
[5]Department of Electrical and Electronic Engineering, University of Cagliari, Via Castelfidardo, Cagliari 93100, Italy
[6]Department of Electrical and Electronic Engineering, Tokyo Institute of Technology, 2-12-1 Ookayama, Meguro-ku, Tokyo, 152-8552, Japan

**S1. Mathematical model of the microring-based magneto-optic modulator**

A central aspect of our work consists in evaluating the optical phase shift generated by the electrical current in the magneto-optic modulator under investigation. In our device, the electrical current in the micro-strip (the electromagnet) is used to generate a magnetic field that transversely magnetizes the Ce:YIG (Voigt configuration[1]). The presence of this external magnetic field turns on the off-diagonal entries of the Ce:YIG permittivity tensor[2], causing the change of the effective index (as well as the phase) of the optical mode in the waveguide. In the reference frame shown in Fig. 1(b) of the manuscript, the permittivity tensor of the Ce:YIG is

$$\varepsilon = \begin{pmatrix} n^2 & 0 & 0 \\ 0 & n^2 & -in\theta_F \lambda/\pi \\ 0 & in\theta_F \lambda/\pi & n^2 \end{pmatrix} \quad (1)$$

where n is the optical refractive index, λ is the optical wavelength, and $\theta_F$ is the Faraday rotation constant. The value of $\theta_F$ depends on the transverse magnetic field, $H_x$, as

$$\theta_F = \theta_F^0 \tanh\left(\frac{H_x}{H_x^0}\right) \quad (2)$$

where $\theta_F^0 = -4500/cm$ and $H_x^0 = 2.4$ mT at room temperature[3]. The Faraday rotation constant varies linearly for small values of $H_x$, and saturates to $\theta_F^0$ when $H_x \geq 2H_x^0$.

The forward (clockwise) and the backward (counter-clockwise) propagating waves exhibit different phase velocities due to the magnetic-induced anisotropy of the Ce:YIG (magneto-optic effect). The variation of the mode effective index compared to the no-current case can be computed using the perturbative approach[4]

$$\Delta n_{eff} = \frac{1}{P_z} \frac{\varepsilon_0 c\, n\, \lambda}{2\pi} \mathbb{Re}\left\{\iint i\, \theta_F(x,y) E_y E_z^*\, dxdy\right\} \quad (3)$$

where $\varepsilon_0$ is the vacuum permittivity, $c$ is the speed of light, ($E_x$,$E_y$,$E_z$) are the electric field components of the optical mode, and $P_z$ is the active power of the optical mode in the propagating direction. In the integral we highlight the dependence of $\theta_F$ on the position, because its value is non-zero only inside the Ce:YIG and it depends on the local magnetic field, as described by Eq. (2)

To evaluate the integral in Eq. (3), we simulate the magnetic field distribution generated by the electrical current in the electromagnet using COMSOL Multiphysics, and we calculate the optical mode profile in the silicon/Ce:YIG waveguide using an electromagnetic mode solver developed in-house[5]. Combining those results, the effective index variation with respect to the electrical current is shown in Fig. 1S. For the sake of completeness, the phase variation per unit length is also shown in the same plot (right-hand y-axis). Figure 1S indicates that controlling the direction and the amplitude of the current in the electromagnet can be effectively used to modulate the effective refractive index of the optical mode underneath, and therefore the resonance of the microring resonator.

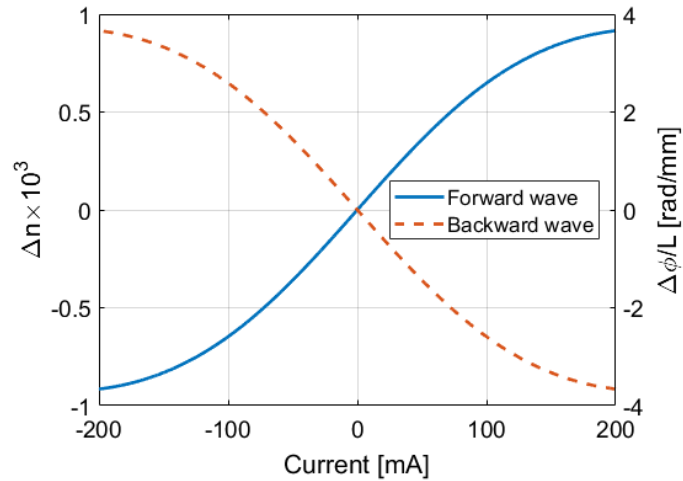

*Figure 1S. The effective index variation (left-hand y-axis) with respect to the current in the electromagnet (x-axis). The corresponding phase variation per unit left is also shown (right-hand y-axis).*

## S2. Energy per bit

The aim of this section is to evaluate the average energy per bit consumed in the magneto-optic modulator. The equivalent circuit of the modulator is shown in Figure 2S, where $R_g$ is the resistance of the modulating circuit, $i_g$ is the current of the generator, $i$ is the current in the coil, $R$ and $L$ are the resistance and the inductance of the coil, respectively. Here, the inductor $L$ describes the electromagnetic energy stored in the circuit, and the resistance $R$ is used to model all the loss in the device, including both direct current (DC) and radio frequency (RF) contributions. The value of $R$ is set mainly by the resistance of the gold film, but it also includes the magnetic loss in the Ce:YIG when the current is alternating.

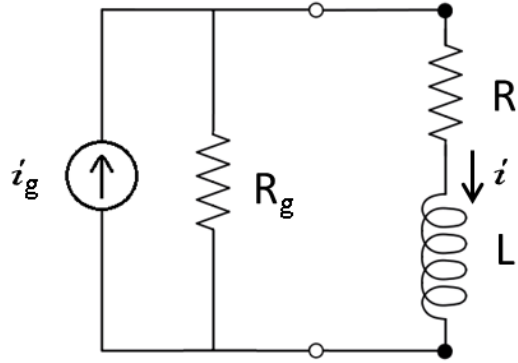

Figure 2S. Equivalent electrical circuit of the proposed modulator.

To determine the energy consumption per bit, the electrical current in the modulator must be computed. Since the modulation is a time varying signal, the equivalent electrical circuit is analyzed in the time domain. The differential equation that relates the current in the electromagnet, i(t), to the current of the generator, $i_g$(t), is the following

$$L\frac{d}{dt}i + (R + R_g)i = R_g i_g \qquad (4)$$

where $i_g$(t) is the input stimulus and i(t) is the output variable. When the current of the generator switches instantaneously from 0 to $I_g$ and the initial current in the electromagnet is 0, the current in the electromagnet changes as

$$i(t) = I\left(1 - e^{-\frac{t}{\tau}}\right)[H(t) - H(t - T_b)] \qquad (5)$$

where *H(t)* is the Heaviside function, and

$$\tau = \frac{L}{R + R_g} \quad \text{and} \quad I = \frac{R_g I_g}{R + R_g} \qquad (6)$$

After the bit interval T<sub>b</sub>, the current of the generator is switched off, and the current in the electromagnet decay exponentially

$$i(t) = I_0 e^{-(t-T_b)/\tau} H(t - T_b) \qquad (7)$$

where

$$I_0 = I\left(1 - e^{-\frac{T_b}{\tau}}\right) \qquad (8)$$

The normalized current in the generator and in the modulator are shown in Figure 3S, in the case of $T_b = 10\,\tau$.

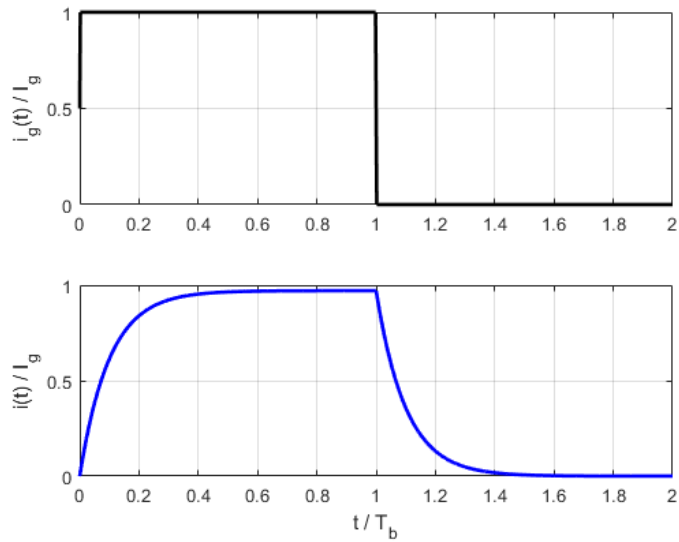

*Figure 3S. Current in the generator and in the modulator. The current values are normalized by the maximum current amplitude of the generator, $I_g$.*

Computing the energy is now straightforward, by integrating the dissipated power over time

$$E_b = \int_0^\infty Ri^2(t)dt + \int_0^\infty R_g[i_g(t) - i(t)]^2 dt \qquad (9)$$

The first term is the energy dissipated in the device, while the second term is the energy dissipated in the resistor of the driving circuit. In the interval [0, $T_b$], the two integrals are

$$\int_0^{T_b} Ri^2(t)dt = RI^2\left[T_b + 2\tau\left(e^{-\frac{T_b}{\tau}} - 1\right) + \frac{\tau}{2}\left(1 - e^{-\frac{2T_b}{\tau}}\right)\right] \qquad (10)$$

$$\int_0^{T_b} R_g[i_g(t) - i(t)]^2 dt = R_g \left[ T_b(I - I_g)^2 + 2\tau \left(e^{-\frac{T_b}{\tau}} - 1\right)I(I - I_g) + \frac{\tau}{2}\left(1 - e^{-\frac{2T_b}{\tau}}\right)I^2 \right] \quad (11)$$

After $T_b$, the current $i_g=0$ and the dissipated energy in the interval [$T_b$, ∞] is

$$\int_{T_b}^{\infty} (R + R_g)i^2(t)dt = \frac{1}{2}(R + R_g)I_0^2 \tau = (R + R_g)I^2 \frac{\tau}{2}\left(1 - e^{-\frac{T_b}{\tau}}\right)^2 \quad (12)$$

The previous terms can be simplified when $T_b$>>τ and R<<$R_g$, indeed

$$\int_0^{T_b} Ri^2(t)dt \cong RI^2 T_b \quad (13)$$

$$\int_0^{T_b} R_g[i_g(t) - i(t)]^2 dt \cong \frac{1}{2}R_g I^2 \tau \cong \frac{1}{2}LI^2 \quad (14)$$

$$\int_{T_b}^{\infty} (R + R_g)i^2(t)dt \cong \frac{1}{2}LI^2 \quad (15)$$

Summing all the contributions, the energy dissipated by a single bit (Figure 3S) is

$$E_b = RI^2 T_b + LI^2 \quad (16)$$

The first term is the energy dissipated in the resistor, while $LI^2$ is the total energy dissipated when the inductor is charged and discharged (1 cycle). The inductor is charged (rising transition) when the current is switched from 0 to I, and it is discharged (falling transition) when the current is switched from I to 0. Each transition dissipates an energy equal to $\Delta E_b = LI^2/2$.

In a bit stream like the one shown in Figure 4S, the current switches between +I and -I and each transition dissipates an energy equal to $\Delta E_b = LI^2$ (discharge and charge of the inductor). The rise and fall transitions appear only when a "1" is followed by a "0" or when a "0" is followed by a "1", so only half of the times in a random sequence. As a result, the average energy per bit previously computed is modified as follow

$$\overline{E_b} = \frac{RI^2}{r_b} + \frac{1}{2}LI^2 \tag{17}$$

where $r_b = 1/T_b$ is the bit rate.

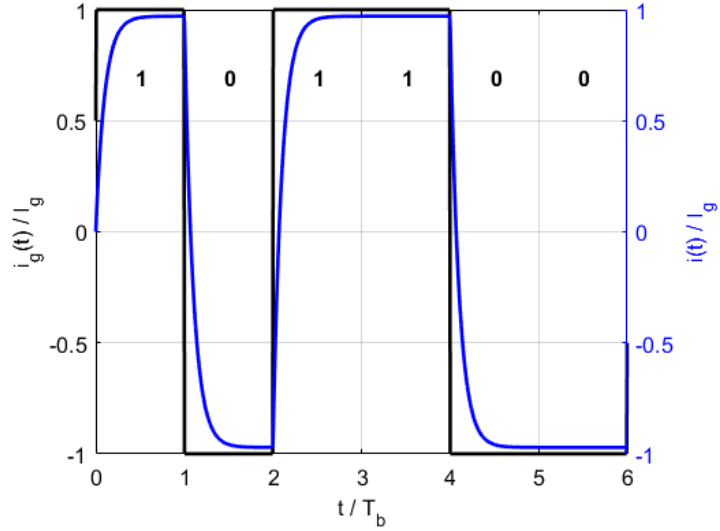

*Figure 4S. Bitstream. Driving signal ($i_g$) and modulated current (i) in the load.*

## S3. Measurements

The electrical and optical characterization of the device is performed in a cryogenic probe station (Montana Cryostation s200) where the temperature is controlled between 4 K and 300 K. The probe station is equipped with one electrical probe for both DC and RF measurements, and two lensed fibers for coupling light to the on-chip waveguides (spot diameter 2.5 μm). The positions of the electrical probe and the lensed fibers are finely regulated via the XYZ piezoelectric nano-positioners (Attocube).

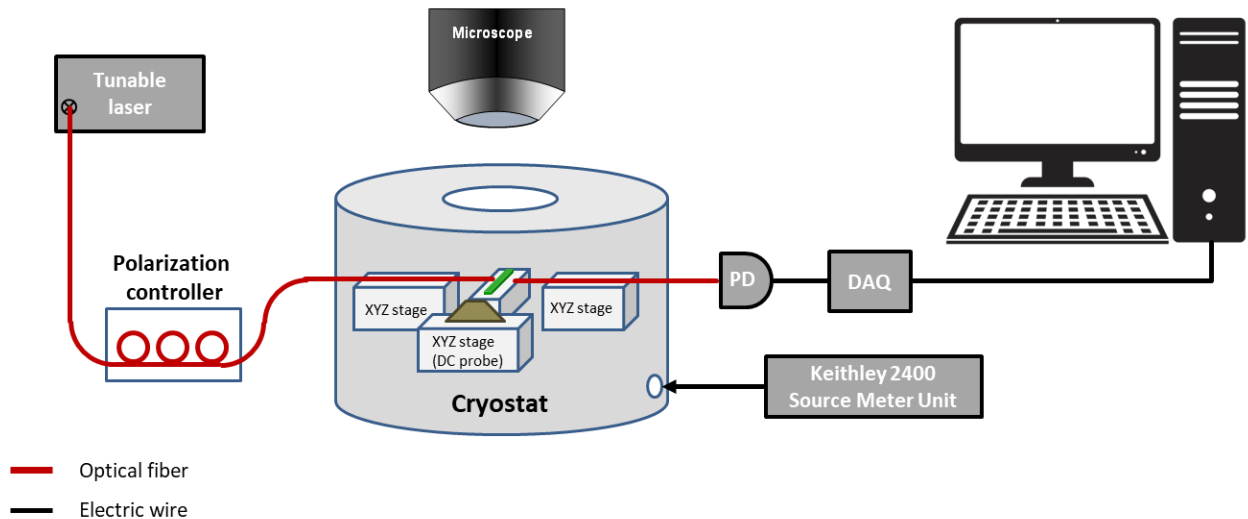

*Figure 5S. Measurement set-up for measuring the optical transmission spectra when different DC input stimuli are applied. The DC signals are generated by a source measure unit (Keithley 2400) which is connected to the device through a DC probe.*

The optical transmission spectra of the magneto-optic modulator are measured in the set up schematically shown in Figure 5S. A continuously tunable external cavity diode laser (Santec) is used to sweep the wavelength of the input light between 1480 nm and 1580 nm. The output signal is collected in a photodetector (PD) and acquired to a desktop computer using a data acquisition board (DAQ). A polarization controller, external to the cryogenic probe station, is used to maximize the transverse magnetic (TM) polarized mode launched into the device. The current in the electromagnet is supplied from a source meter unit (Keithley 2400).

### S3.1. Resistance and inductance characterization

To evaluate the energy-per-bit of the magneto-optic modulator, the resistance and the inductance of the device are measured between 4K and room temperature.

The measured and simulated values of the DC resistance are shown in Figure 6S, where a coil with a diameter of 70 µm is considered. The resistance, R, is measured with a 4-probe 6-digit Digital Multimeter (DMM). Its value is 1.43 Ω at 300 K and drops to ~350 mΩ at low temperatures. The calculation of the resistance with respect to the temperature is performed with COMSOL Multiphysics using the a conductivity for gold of 22.7 nΩ·m at 300 K, 4.81 nΩ·m at 77 K, and 0.22 nΩ·m at 4 K,, as reported in the literature[6]. While the calculated and measured resistance are very close in the range 100 K-300 K, the two curves diverge for below 100 K. This discrepancy is caused by impurities or grain boundaries in the gold layer, which limit the conductance of the coil.

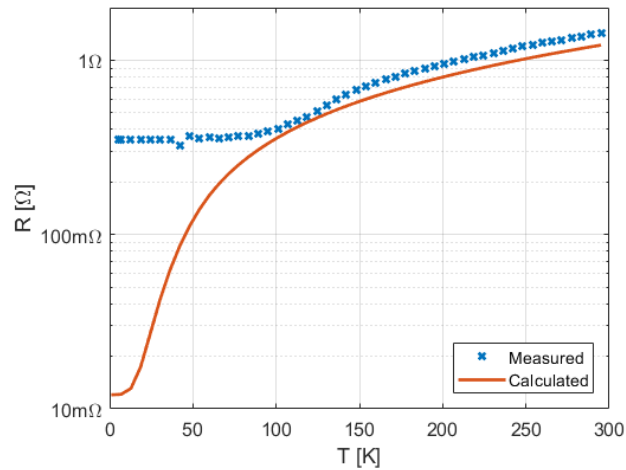

Figure 6S. Measured and calculated resistance at direct current (DC) of the magneto-optic modulator. The relative errors between the simulated and the measured resistance is below 15% in the range 100 K-300 K.

The behavior of the resistance and the inductance as functions of the frequency are derived from the measurements of the electrical reflection coefficient, $S_{11}$, knowing that $S_{11}= (R+j\omega L- Z_0) / (R+j\omega L+ Z_0)$, where $Z_0 = 50\ \Omega$ is the characteristic impedance of the transmission line, and $\omega$ is the angular frequency. Coefficient $S_{11}$ s measured with a vector network analyzer at 4 K and 300 K, after the setup was calibrated (including the probe tips) at both temperatures. The amplitude and the phase of $S_{11}$ are shown in Figure 7S for four different devices, where the coil has a diameter d=40 μm, d=50 μm, d=60 μm, and d=70 μm, respectively. The amplitude of $S_{11}$ at 4 K shows some small oscillation (<0.2dB) due to unavoidable electrical reflections.

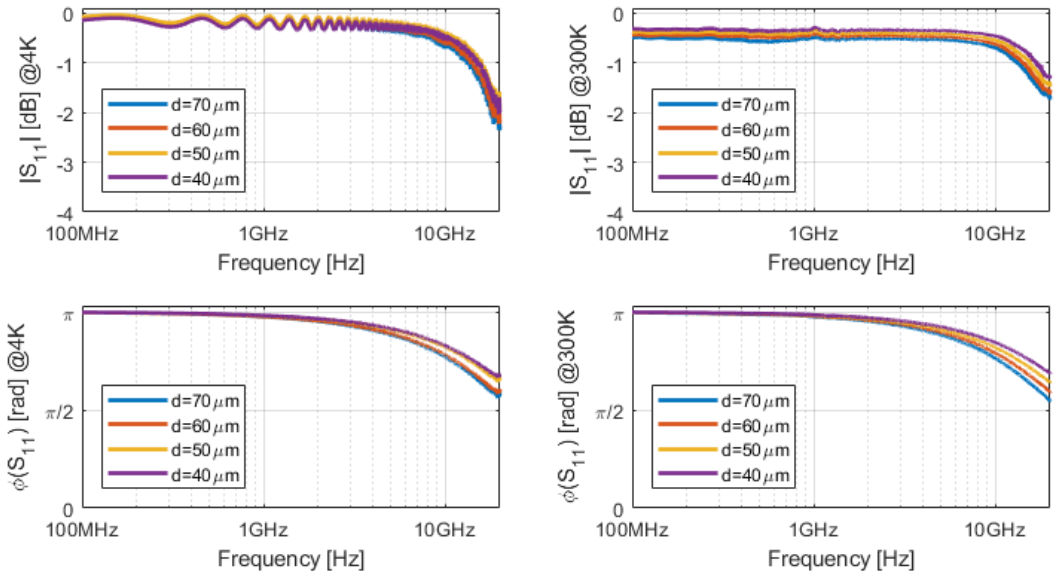

Figure 7S. Amplitude (top) and phase (bottom) of the electrical reflection coefficient $S_{11}$ at 4 K (left) and 300 K (right).

The value of R and L are extracted from the measured $S_{11}$ and shown in Figure 8S. Since the value of R is quite small compared to characteristic impedance of a transmission line, the small oscillations on the measurements of $S_{11}$ at 4 K affect the accurate estimation of the resistance. However, we can observe that the average value of R extracted from $S_{11}$ at low frequency is consistent with the DC measurements shown in Fig. 6S. On the other hand, we find that the value of L does not change significantly with the frequency. Those two observations allows us to affirm that there is no significant magnetic dissipation in the Ce:YIG. If there was, that would show up as an increase in R and a decrease in L with frequency as the real part of the Ce:YIG permeability drops[7]. The increase of R at frequencies larger than 10 GHz is due to the skin effect in the electromagnet. The skin depth of gold is about 0.75 μm at 10GHz, which is comparable to the thickness of the wire of 1.5 μm. It is also worth noting that while R drops when the temperature decreases, the value of L does not change significantly with temperature.

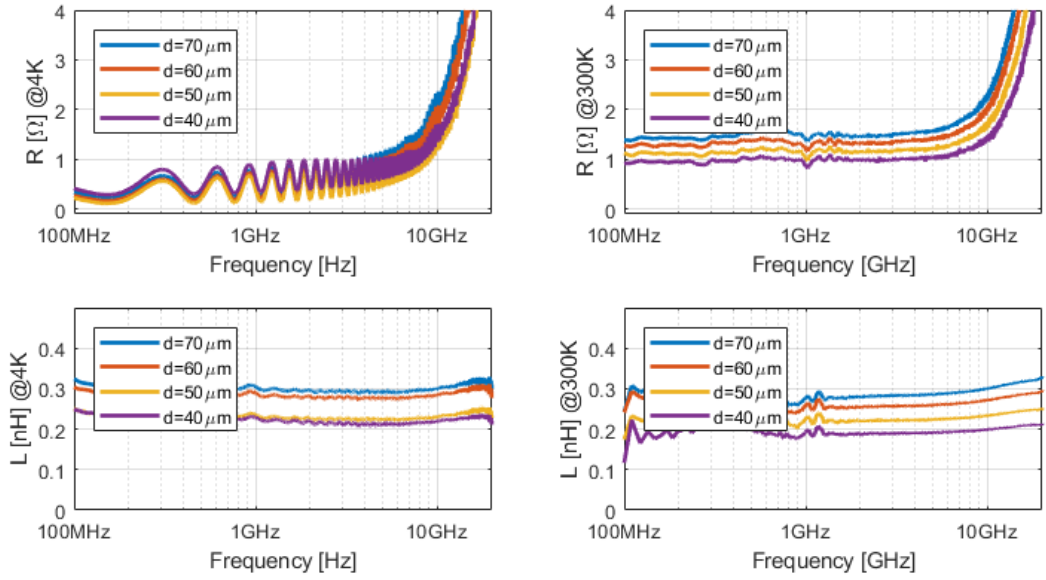

*Figure 8S. Resistance and inductance of the magneto-optic modulator at 4 K (left) and 300 K (right).*

The calculated and measured values of R and L are compared in the table below, showing excellent agreement for L in the range 4 K-300 K, while the simulated values of R match the measurements only in the range 100 K-300 K, as previously explained. The calculations account for the electrical and magnetic characteristics of all the materials, which are reported in the supplementary materials of *D. Huang et al.*[8].

|   | Simulation 4K | Measurement 4K | Simulation 77K | Measurement 77K | Simulation 300K | Measurement 300 K |
|---|---|---|---|---|---|---|
| R | 12 mΩ | **347 mΩ** | 251 mΩ | **365 mΩ** | 1.22 Ω | **1.43 Ω** |
| L | 0.250 nH | **0.296 nH** | 0.250 nH | - | 0.250 nH | **0.288 nH** |

**S3.2. Insertion loss and propagation loss**

The optical loss in the magneto-optic modulator is measured by comparing our device to a standard straight silicon waveguide with the same dimensions but with a silica cladding instead of the garnet. The length of the bonded Ce:YIG (3.5mm) is much longer than a single device (70 μm diameter), such that eight modulators, including the electrical pads, can be fitted in the mask, as shown in Figure 9S.

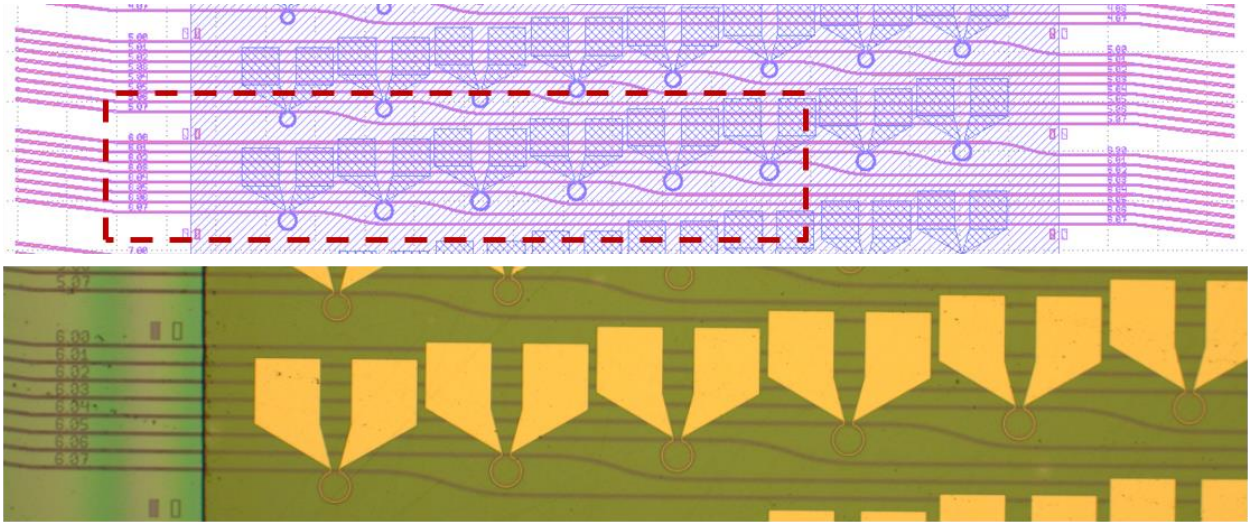

*Figure 9S. Mask layout (top) and fabricated chip (bottom).*

The measured insertion loss is 10 dB, which consists of propagation loss and scattering from the bonded chip interfaces. To distinguish between the two contributions, we simulate the scattering loss at the interfaces between the waveguide with a silica cladding and the waveguide with bonded Ce:YIG. The simulated scattering loss per interface is about 0.6 dB, as shown in Figure 10S, and we attribute the remaining 8.8 dB mainly to propagation loss. Therefore, the excess loss of a single modulator is (1.2+8.8/8) = 2.3dB.

The excess loss can be further reduced by using silicon nitride cladding (n = 1.99) in place of silicon dioxide (n = 1.45) due to a smaller refractive index contrast with the Ce:YIG (n = 2.22), as shown Figure 10S. With this approach, the excess loss can be reduced below 1.5 dB.

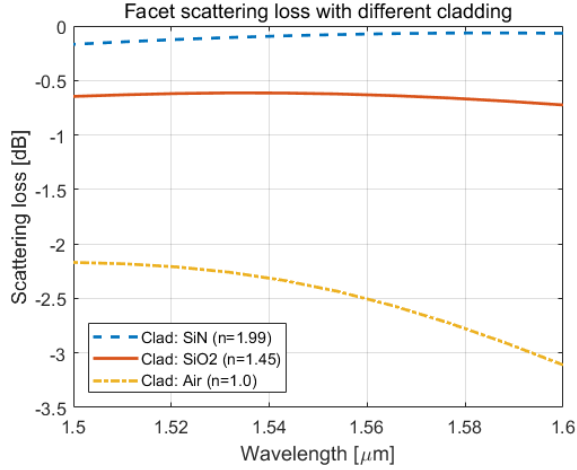

*Figure 10S. Scattering at the interface between waveguide and bonded Ce:YIG when different cladding are used for the silicon waveguide.*

**S3.3. Power coupling coefficient and loss per roundtrip**

In the manuscript we show that the highest extinction ratio (ER) is achieved when the critical coupling condition is met, while the energy consumption per bit ($E_b$) is lower when the modulator operates in the under-coupled regime. In this section, we show how we extracted the power coupling coefficient, $K$, and the roundtrip loss, $\gamma$, from the optical spectrum of the microring modulator. For the sake of clarity, let us recall that the power coupling coefficient, $K$, is the fraction of optical power in the straight waveguide that enters the microring resonator (or vice versa), and the roundtrip loss, $\gamma$, is the power lost when the light travels once along the ring[9]. If the propagation loss per unit length is $\alpha$, we have $\gamma = 2\pi r \cdot \alpha$ where $r$ is the microring radius.

The values of K and $\gamma$ are related to the full width at half maximum (FWHM) and the extinction ratio (ER) of the microring response through the following equations[10]:

$$FWHM = \frac{(1 - t \cdot a)\lambda^2}{2\pi^2 r \cdot n_g \cdot \sqrt{t \cdot a}} \tag{18}$$

$$ER = \frac{(t+a)^2}{(t-a)^2} \cdot \frac{(1-t\cdot a)^2}{(1+t\cdot a)^2} \qquad (19)$$

$$a = \exp(-\gamma/2) \qquad (20)$$

$$t = \sqrt{1-K} \qquad (21)$$

An over-coupled (OC) and an under-coupled (UC) resonator can give the same values for FWHM and ER, so we use Figure 11S to estimate the values of K and $\gamma$. In this figure, the contours of the FWHM and the ER are shown in the ($\gamma$, K)-space. On the same chart, we show the experimental values for FWHM and ER of the resonator at λ=1500 nm and λ=1550 nm. The circle markers identify the pairs of FWHM and ER assuming the resonator is in the OC condition and, correspondingly, the square markers indicate the pairs of FWHM and ER if the resonator operates in the UC condition. It is worth noting that the contour plots are symmetric with respect to the line K= $\gamma$ (i.e., critical coupling condition).

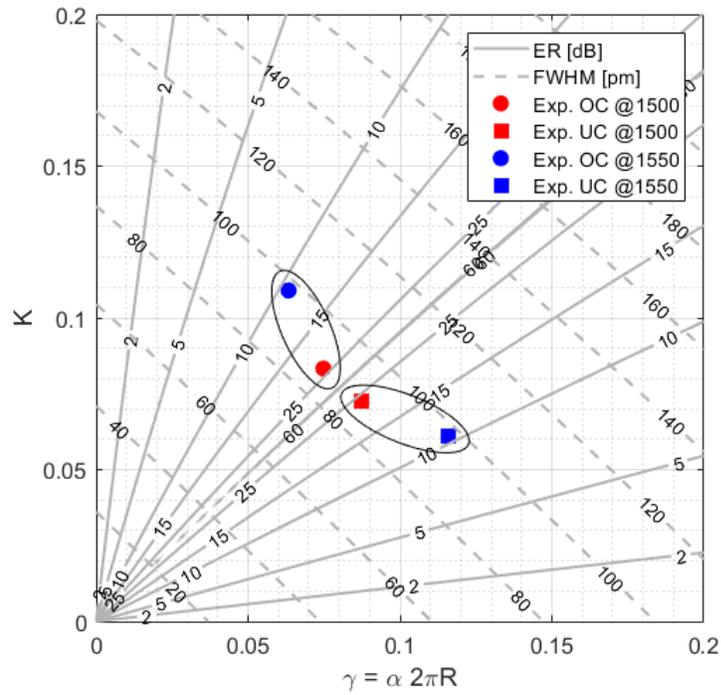

Figure 11S. Contour map of FWHM and ER as a function of $\gamma$ and K. The experimental results are shown assuming the resonator is in the over-coupled and the under-coupled condition, respectively.

To solve the ambiguity and extract the values for $\gamma$ and K, we consider two operating wavelengths, λ=1500 nm and λ=1550 nm and make the following observations:

1. the resonator is either in the OC or UC condition, and this condition is the same for both wavelengths;
2. the propagation loss and the roundtrip loss are expected to be very similar at 1500 nm and 1550 nm. Indeed, the mode profile does not change significantly between these two wavelength as demonstrated by the confinement factor in the Ce:YIG shown in Figure 12S(a);
3. the simulated coupling coefficient K varies significantly between the two wavelengths, as shown in Figure 12S(b).

From Figure 11S, we can observe that the two OC points (circles) have similar loss but different coupling coefficient. In contrast, the two UC points (squared) have similar coupling coefficient but different loss. From point 2 and 3, we can conclude that the resonator operates in the over-coupled regime at both wavelengths.

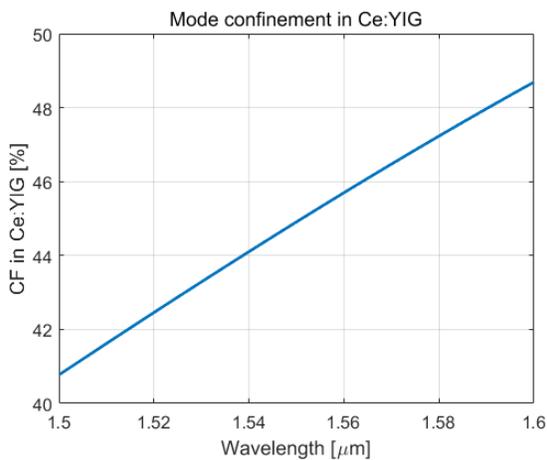
(a)

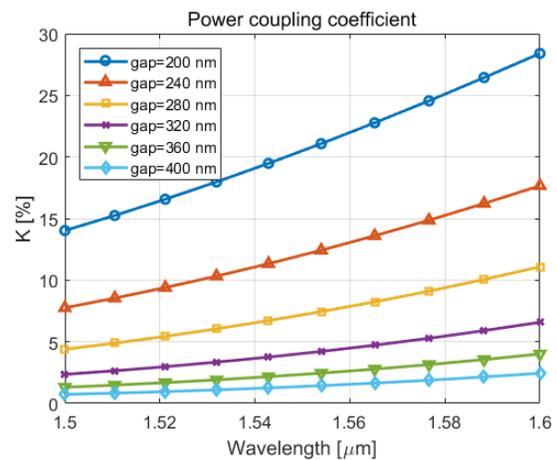
(b)

*Figure 12S. Simulated confinement factor (CF) in Ce:YIG and power coupling coefficient in the magneto-optic microring modulator.*

## S3.4 Magnetic response of Ce:YIG

Figures 13S(a)-(d) show the in-plane magnetization hysteresis loops of Ce:YIG at temperatures of 4 K, 77 K, 173 K and 300 K, respectively, where the linear paramagnetic component from the SGGG substrate was subtracted. The magnetization loops were measured with a superconducting quantum interference device (SQUID: MPMS-XL, QUANTUM DESIGN). The result shows that the loop opens as the temperature is lowered, with a higher residual magnetization and stronger coercivity. The magnetic anisotropy of ferrimagnetic material such as Ce:YIG typically becomes high at a low temperature because the thermal energy at higher temperature breaks magnetic ordering, reducing the magnetic anisotropy.

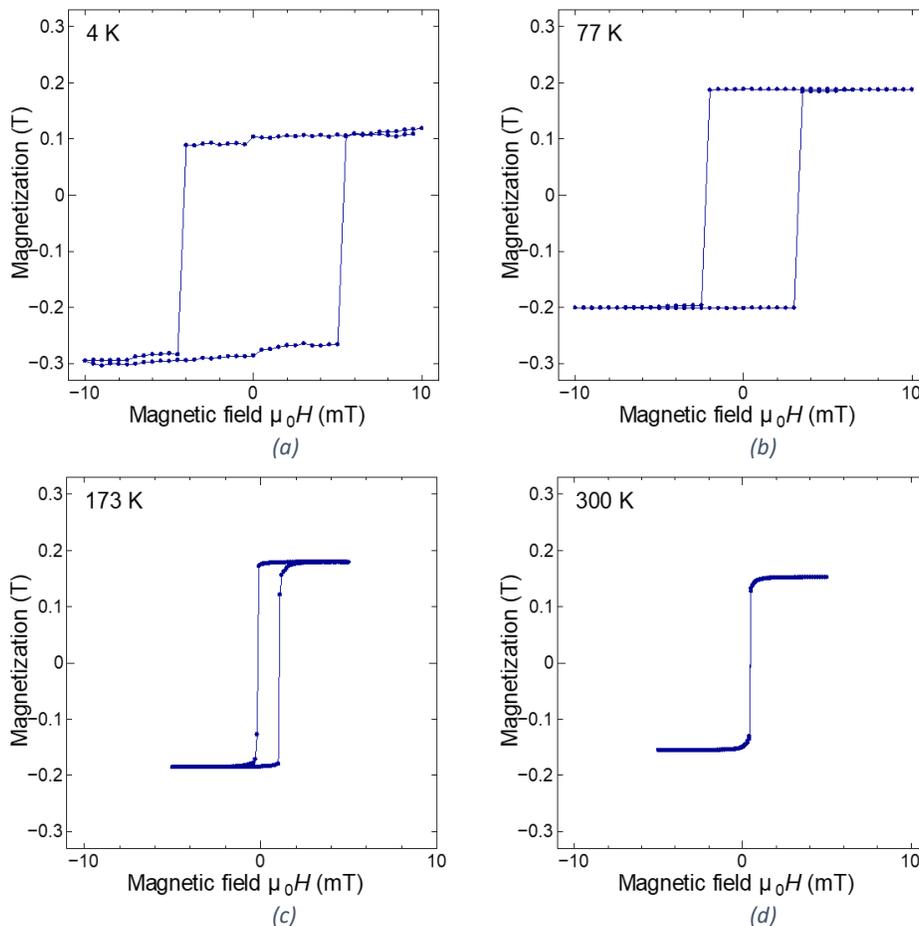

Figure 13S. Magnetic hysteresis loops at (a) 4 K, (b) 77 K, (c) 173 K, and (d) 300 K measured by SQUID. As a comparison, the magnetic field generated by the electromagnet when I = 110mA at 5 μm (i.e., CeYIG/silicon interface) is about 15 Oe (1.5 mT).

When the current in the electromagnet is 110 mA, the in-plane radial magnetic field is about 1.5 mT that is below the magnetic saturation of the Ce:YIG (see the measurements in Fig. 13S). This results in a linear MO split with respect to the applied current, as confirmed by the measurements shown in Figure 3(c) of the manuscript. The value of the residual magnetization is evaluated between 4 K and 77 K by measuring the spectrum in the clock-wise (CW) and in the counter-clock wise (CCW) directions, when no current is applied in the electromagnet. The measured spectra are shown in Figure 2(a) in the manuscript.

The resonance of the microring red-shifts when the temperature rises, as shown in Figure 14S(a). For both CW and CCW resonances, the thermo-optic coefficient is extracted from the shift of the resonance according to

$$\frac{dn_{eff}}{dT} = \left[\frac{dn_{eff}}{dn_{Si}}\frac{dn_{Si}}{dT} + \frac{dn_{eff}}{dn_{SiO2}}\frac{dn_{SiO2}}{dT} + \frac{dn_{eff}}{dn_{CeYIG}}\frac{dn_{CeYIG}}{dT} + \frac{dn_{eff}}{dn_{Air}}\frac{dn_{Air}}{dT}\right] = \frac{\lambda}{n_g}\frac{d\lambda}{dT} \quad (22)$$

where $dn_i/dT$ is the thermo-optic coefficient of the material (i=Si, SiO$_2$, Ce:YIG and vacuum), and $dn_{eff}/dn_i$ is a function of the waveguide geometry. In the equation above, we highlight the contribution of the four materials for the sake of clarity.

We observe a constant offset between the CW and the CCW resonance for all temperatures between 4 K and 77 K, as shown in Figure 14S(b). The average offset is only 3.74 pm, which suggests the presence of a very small residual in-plane magnetization for the Ce:YIG. The lack of a permanent magnetization is explained observing that the size of the magnetic domains in the Ce:YIG is comparable with the size of the 70 μm-diameter microring resonator, as show in Figure 14S(c). Because the microring resonator has too small a radius to pin the residual magnetization in a radial direction, the domain easily randomizes when no current is applied in the electromagnet.

As shown in Figure 3(c) of the manuscript, the MO split magnitude is lower at 4 K and 77 K than at 300 K, which goes counter to what we would expect based on the SQUID measurement, where the saturation magnetization increases at lower temperatures. At this point we observed: (i) the residual magnetization becomes negligible under weak applied magnetic field, (ii) the MO response is linear in current because Ce:YIG is magnetized only in the area underneath the gold wire. These two considerations allow us to conclude that the MO split is proportional to the coercive force, which is stronger at low temperature. Therefore, a higher current is needed at 4 K and 77 K to obtain the same MO effect that we see at 300 K.

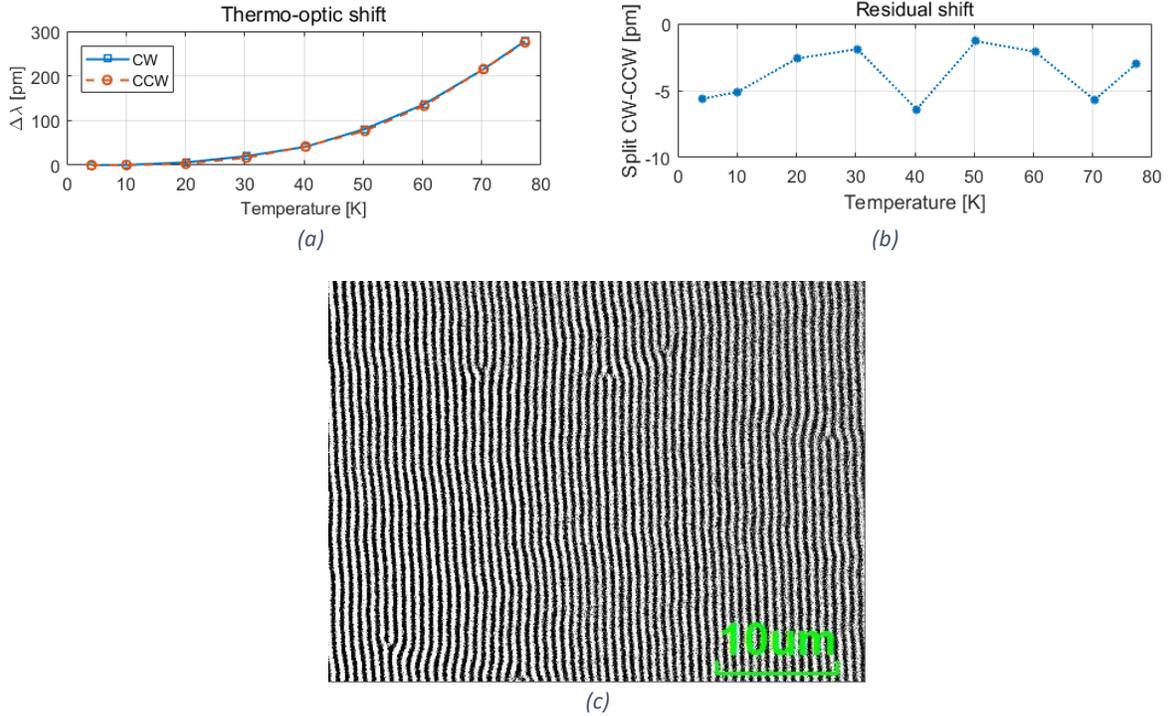

Figure 14S. (a) Measured resonance shift with respect to the spectra at 4 K, (b) resonance split between clock-wise and counter-clock-wise resonance. (c) Magnetic domain of Ce:YIG film at 300 K measured by MO Kerr microscope.

**S3.5 High speed measurement set-up**

The high-speed measurement set-up is schematically shown in Figure 15S, where the continuous lines refer to the set-up for measuring modulation bandwidth, whereas the dashed lines denote the set-up for eye diagram measurements.

The modulation bandwidth is measured using a vector network analyzer (VNA), which generates an RF signal that is swept in frequency, and reads back the amplitude and phase of the signal at each stimulus frequency on a separate port. A benchtop tunable laser source generates the 1550 nm optical carrier to be modulated. The optical signal from the output of the modulator is received by a photodetector connected to the receiving port of the VNA. The optical amplifier and the tunable optical filter are not used for this measurement. The device has been tested in the frequency range from 100 MHz to 20 GHz, and the experimental results are shown in Figure 4(a) in the manuscript.

For measuring the eye diagrams shown in Figure 4(d) of the manuscript, a $2^9$-1 pseudorandom binary sequence (PRBS) is generated by a stand-alone Bit Error Rate Tester (BERT, Anritsu) and amplified by a driver amplifier[11], which can output a maximum amplitude of 8V peak-to-peak. The signal is reduced by 3 dB after the driver amplifier, to avoid the risk of component damage. In this configuration, we estimated a maximum current of 110 mA through the input coil of the device for an output amplitude of 8 $V_{pp}$ from the driver. Using this value and the measured resistances, we estimate the maximum dissipated power in the modulator at all temperatures.

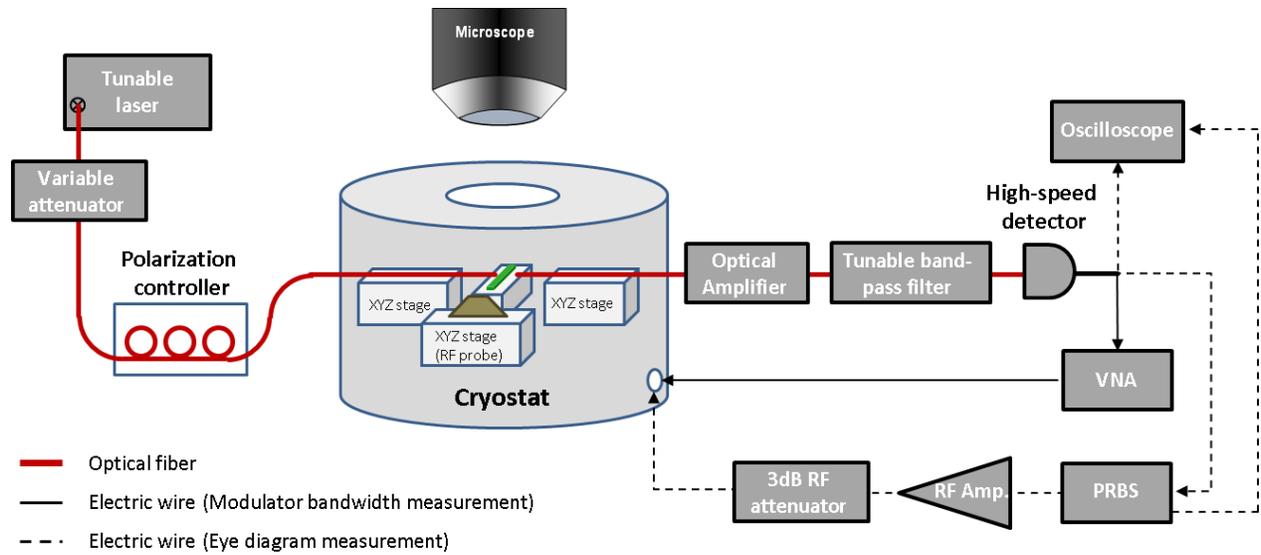

*Figure 15S. Measurement set-up for the bandwidth and the eye diagram measurements.*

**S4. Thickness optimization of gadolinium gallium garnet substrate**

The distance between the electromagnet and the optical mode is one of the most critical parameters of the magneto-optic modulator, since it determines how much electrical current is needed to magnetize the Ce:YIG as well as how much power is required. Although having a small distance is desirable, it requires the SGGG to be mechanically polished with high precision after the flip chip bonding.

To evaluate the benefit of reducing the SGGG substrate, we calculated the magnetic field at the silicon/Ce:YIG interface as a function of the SGGG thickness for different values of the electrical current, as show in Figure 16S. For the device investigated in this work, the SGGG is 5 µm thick and a current of 110 mA is required to achieve largest magneto-optical modulation. Reducing the SGGG layer down to 1 µm benefits from halving the electrical current so saving up to 75% of energy per bit compared to the present device. This result shows the energy consumption can be greatly reduced by optimizing the fabrication process.

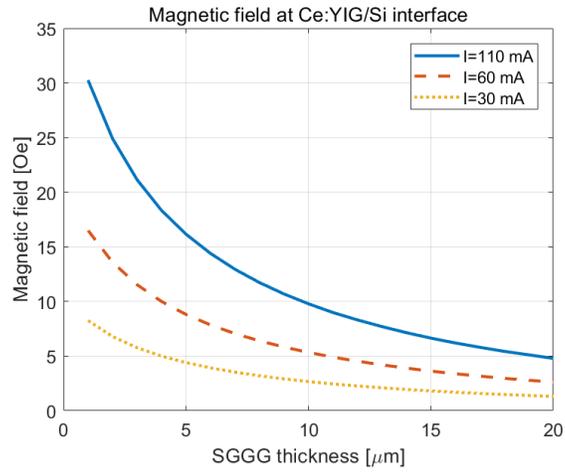

*Figure 16S. Magnetic field amplitude at the Ce:YIG/Silicon waveguide interface as a function of the substituted gadolinium gallium garnet substrate thickness.*